# Replicating the flyby sampling of salty ocean world ice grains using impact ionization mass spectrometry

K. Marshall Seaton,[1]*† Bryana L. Henderson,[1]* Sascha Kempf,[2] Sarah E. Waller,[1] Morgan E. C. Miller,[1] Paul D. Asimow,[3] Morgan L. Cable[1]

[1]NASA Jet Propulsion Laboratory, California Institute of Technology, Pasadena, CA.
[2]Laboratory for Atmospheric and Space Physics, University of Colorado, Boulder, CO.
[3]Division of Geological and Planetary Sciences, California Institute of Technology, Pasadena, CA.

*Corresponding authors: K. Marshall Seaton, Bryana L. Henderson
Emails: marshall.seaton@lasp.colorado.edu; bryana.l.henderson@jpl.nasa.gov

†Present address: Laboratory for Atmospheric and Space Physics, University of Colorado, Boulder, CO.

## Abstract

The Europa Clipper mission will arrive at the Jovian system in 2030 and analyze ice grains sourced from the icy material on its surface using impact mass spectrometry, which will provide key constraints on Europa's chemical composition and habitability. However, deriving quantitative compositional information from spaceborne impact mass spectra of ice grains has historically proven difficult due to the confounding effects of composition and impact velocity, coupled with difficulties in accelerating ice grains to spacecraft velocities under analogous sampling conditions. Using a novel hypervelocity ice grain acceleration and impact mass spectrometry method, we quantify the degree to which the mass spectra of NaCl-rich ice grains are influenced by chemical composition and impact velocity variations within the flyby velocity ranges planned for the Europa Clipper mission. These results suggest that high-fidelity studies quantifying composition and velocity-related effects in impact mass spectra may be necessary to accurately interpret data collected at Europa and other ocean worlds in the future.

## Introduction

Impact ionization mass spectrometers, colloquially referred to as dust analyzers, have enabled a myriad of discoveries throughout the solar system spanning several decades of planetary science mission efforts, allowing scientists to study the mass, charge, directionality, and chemical composition of spaceborne dust grains (*1–5*). In contrast to other mass spectrometers included in mission payloads, which typically ionize samples using controlled energies (*6–9*), impact mass spectrometers ionize spaceborne dust grains through hypervelocity impact of the particle onto a metal target within the instrument during a spacecraft encounter. At high enough spacecraft velocities (>1 km/s), these dust grain impacts provide the kinetic energy necessary to generate an impact plasma containing ions, neutral species, and molecular fragments. The ions generated from these impacts are analyzed using time-of-flight mass spectrometry (TOF-MS), allowing chemical compositional analysis of the sampled material.

Although not originally proposed for the analysis of ice grains specifically, the Cassini Cosmic Dust Analyzer (CDA) instrument clearly demonstrated the advantages of impact ionization mass spectrometers in the analysis of ice grains sourced from ocean worlds during its tour of the Saturnian system. Perhaps one of the most notable of these achievements includes evidence for a salty liquid water ocean as the source of the Enceladus plume through the detection of NaCl and Na-carbonate rich ice grains sourced from its subsurface ocean (*10*) in addition to a plethora of other insights into the habitability and interior chemistry of Enceladus (*11–13*). Drawing upon heritage from the CDA and its predecessors on the Ulysses and Galileo missions, the Surface Dust Analyzer (SUDA) on Europa Clipper will arrive at the Jovian system in 2030 and analyze ice grains originating from Europa's surface, and perhaps from potential plumes (*14*). In addition, SUDA measurements will uniquely enable compositional mapping of Europa's surface via the analysis of material ejected by micrometeorite bombardment (*15*, *16*), providing deeper insight into the chemical composition and provenance of potentially endogenous surface features (Fig. 1). These *in situ* measurements, in combination with complementary spectroscopic and geophysical investigations (*17–19*), will be critical in providing a comprehensive assessment of Europa's habitability.

However, like much of the data collected during planetary science missions, the analysis of spaceborne impact mass spectra is nontrivial. CDA data collected at the Saturnian System qualitatively demonstrated that the appearance and relative signal intensity of cluster species observed in ice grain impact mass spectra are strongly influenced not only by chemical composition, but also impact velocity (*20*). As a result, establishing firm quantitative relationships between any observed mass spectral features and the environments from which they are sourced is challenging. This unique aspect of impact mass spectrometry necessitates the implementation of high-fidelity technologies capable of not only varying ice grain composition, but also impact velocity, to replicate this process in the laboratory and provide the necessary ground-truth measurements with which to interpret spacecraft data collected at ocean worlds.

Previous interpretations of ice grain composition at ocean worlds have been carried out through analogue experiments aimed at replicating the impact process through laser irradiation of a water beam, which has formed the basis for our interpretation of ice grain composition in impact mass spectra thus far (*10*, *12*, *13*). The ions generated through this desorption process were analyzed directly and assumed to be representative of the ions generated through hypervelocity impact (*21*). Although these experiments have been foundational in providing preliminary assessments of ice grain and ocean composition from data collected at Enceladus, it is important to note that these experiments do not replicate the unique physics and chemistry of a hypervelocity impact and therefore may not be fully representative (*22*). However, studies replicating the hypervelocity ice grain sampling process in the laboratory have proven difficult to conduct due to the technical limitations associated with accelerating ice grains to typical spacecraft flyby velocities under planetary flyby sampling conditions (3-5 km/s) (*23*). To address this need, Ulibarri et al. have carried out experiments impacting metal particles onto ice (termed the "inverse experiment"). Although this approach allows access to very high range of impact velocities and replicates the impact process, the degree to which a metal particle impacting ice is analogous to the impact of an ice grain onto metal is currently under investigation (*24*). Burke et al. have also recently developed an ice impact mass spectrometer to support biosignature detection efforts utilizing impact mass spectrometry, which uses electrospray ionization (ESI) to generate charged ice grains which are accelerated to velocities up to 4.2 km/s (*25, 26*). A similar approach was recently adapted by Spesyvyi et al. (*27*). Although these approaches replicate the ice grain impact process at relevant velocities, limitations of ESI-based methods using salt-rich solutions (decreased ionization efficiency, ion yield and spray stability) (*28, 29*) limits their applicability in the study of ice grains containing high salt concentrations like those expected at ocean worlds in our solar system. These technical limitations have prohibited impact studies of hypersaline ice grains thus far, which could affect the assessment of Europa's surface (and therefore interior) composition.

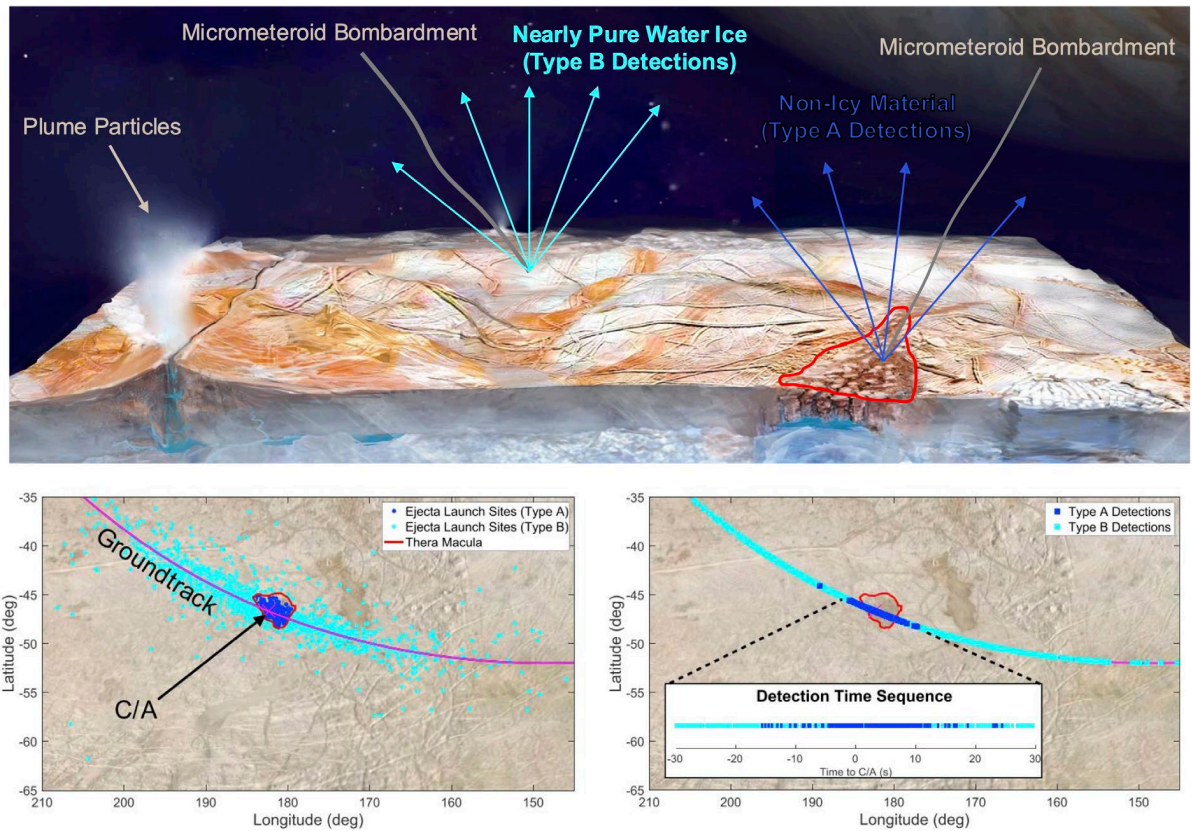

**Figure 1. Analysis of exospheric ice grains at Europa by SUDA.** In addition to analyzing any potential plume material, the Europa Clipper Surface Dust Analyzer (SUDA) investigation will detect (and potentially quantify) chemical species embedded in ice grains directly sourced from Europa's surface by constraining their area of origin. Continuous micrometeoroid bombardment lofts surface material into Europa's exosphere, which will be analyzed directly by SUDA during flyby. Non-icy surface material is expected to be compositionally distinct from the nearly pure water ice material comprising Europa's surface "background" (Top; noted as Type A detections and Type B detections, respectively). Using an ejecta cloud model coupled with SUDA's ability to determine the velocity vector of impacting particles, the area from which each ice grain originated can be determined with a spatial resolution dependent on spacecraft altitude (*15*). A single Monte Carlo simulation for SUDA detections over chaos terrain during the sixth Europa flyby in the 19F23v2 tour design are shown (bottom). The left plot shows the origin of ice grains detected by SUDA; the right plot shows the simulated time sequence for these detections. C/A marks the point of closest approach. Tracing ice grain composition to an area on Europa's surface will provide ground-truth measurements for interpreting remote sensing data and enable direct compositional measurements of Europa's surface geology. Bottom images adapted with permission from ref. (*16*). Image credit: NASA/JPL-Caltech.

To address this need, we have developed an impact mass spectrometry system termed the Hypervelocity Ice grain Impact Validation Experiment (HIIVE). Our method uniquely enables the analysis of hypersaline ice grain impacts collected at spacecraft flyby velocities in the same manner as a spaceborne dust analyzer such as CDA or SUDA. We present impact mass spectra of NaCl-rich ice grains and show that significant differences in the speciation of salt clusters in impact mass spectra not only occur with variations in ice grain composition, but also with changes in impact velocity within the range of planned Europa Clipper flyby velocities. In addition to informing future missions to sample the Enceladus plume (*30*), these studies will be crucial to support the upcoming Europa Clipper mission's SUDA investigation interrogating exospheric ice grains and performing compositional mapping of Europa's surface, where hydrated salts are expected to dominate the composition of non-icy surface material (*31*).

## Results

Our impact mass spectrometry method is described in detail in the Materials and Methods section. Briefly, nm- to µm-sized water droplets are generated directly through the mechanical shock-wave dispersion of a water jet. These droplets freeze rapidly under high vacuum conditions, forming ice grains (*26*, *27*). This process is exceptionally tolerant to high salt concentrations, as the dispersion process is not influenced significantly by the salinity of the solution from which the ice grains are sourced (*21*). Ice grains are accelerated to high velocities through the dispersion process, with a fraction of them on suitable trajectories to encounter (after a delay time inversely proportional to their velocity) a metal impact target. The impact generates a plasma containing neutral and ionized atoms, molecules, and larger fragments. Using pulsed ion optics with precisely defined extraction timing, ions generated through ice grain impacts at a selectable impact velocity are extracted and analyzed using TOF-MS. The impact velocities achieved through this approach (1.9 – 4.5 km/s) overlap with the flyby velocities planned for future ocean world missions and mission concepts (3.0 – 5.0 km/s) (*23*, *30*, *32*), which enables this experimental method to approximate an ocean world flyby ice grain sampling event by a spaceborne dust analyzer.

In this study, we conducted ice grain impact ionization experiments at 3.9 km/s using an Enceladus ocean simulant containing 0.2 M NaCl and 0.1 M $Na_2CO_3$, as well as pure NaCl solutions at concentrations of 0.1 M, 1.0 M, and 3.0 M. To assess the influence of impact velocity on the resulting mass spectra of salt-rich ice grains, we systematically varied the extraction timing (a proxy for velocity) to target impact velocities of 2.4 to 3.6 km/s using a 1.0 M NaCl solution.

### Impact Mass Spectra of NaCl-Rich Ice Grains

*Simulated Enceladus Composition* – Although the vast majority of Saturn's E-ring grains (which are directly sourced from the Enceladus plume) (*33*) consist of pure water ice, previous analyses of larger E-ring ice grains by CDA show that many of the larger grains contain $Na^+$ mass lines (*10*). Of these, approximately 6% of the mass spectra in the analyzed dataset contained strong features that indicate high $Na^+$ and $Cl^-$ content. Estimates of Enceladus ocean composition through the analysis of these salt-rich datasets suggest that the ocean is composed predominantly of NaCl and Na-carbonates, which are estimated to range from 0.05 – 0.2 M and 0.01 – 0.1 M, respectively, based on prior laboratory analogue experiments (*10*). Motivated by these estimates, ice grains generated using an aqueous solution of 0.2 M NaCl and 0.1 M $Na_2CO_3$ were analyzed using our laboratory impact ionization mass spectrometry method. Our data approximately reproduces both the presence and relative ion counts of characteristic mass lines present in Enceladus plume ice grain mass spectra as measured by Cassini (Fig. 2). We note that, although our data show a strong resemblance to the CDA data, our salt concentrations were not intended to provide a quantitative match. Examining such relationships in detail in the context of the CDA data collected at Enceladus is out of the scope of this work and would require further study.

The composition of salt-rich E-ring ice grains sourced from Enceladus was previously assessed by co-adding the CDA mass spectra of many individual ice grains (*10*). The resulting mass spectra are characterized by prominent $Na^+$, $(NaOH)_nNa^+$, and $(NaCl)_nNa^+$ cluster ions, and contain minor signal contributions from $K^+$, $(H_2O)Na^+$, $(NaOH)(NaCl)Na^+$, and $(Na_2CO_3)Na^+$ clusters (*10*). The $Na^+$, $(NaOH)Na^+$, $(NaCl)Na^+$ ions are easily identifiable in a single-grain impact mass spectrum collected by the CDA at Enceladus (Fig. 2, top), with the relatively low mass resolution of the CDA resulting in broader (albeit still distinguishable) features at higher masses including the $(NaOH)_2Na^+$ and $(NaOH)NaCl)Na^+$ clusters. The $(Na_2CO_3)Na^+$ cluster was not clearly identified in the CDA mass spectrum shown here, however this ion is typically very low in abundance and difficult to identify without spectral averaging or co-adding given the limited mass resolution of the CDA. In our data, we observe peaks corresponding to the $Na^+$, $(NaOH)_nNa^+$, $(NaCl)_nNa^+$ $(H_2O)_nNa^+$, $(NaOH)(NaCl)Na^+$, and $(Na_2CO_3)Na^+$ clusters, all of which are expected for NaCl- and Na-carbonate rich ice grain mass spectra based on data collected at Enceladus by the CDA (*10*). The peaks at 41 and 59 m/z observed in our data, which includes signal contribution from the $(H_2O)Na^+$ ion (*SI Appendix*, Fig. S1) and the $(H_2O)_2Na^+$ cluster ion, respectively, are not observed in the individual CDA ice grain mass spectrum shown here (CDA impact event ID 180283 retrieved from the PDS) and are very low abundance in published data representing bulk Enceladus ocean composition (*10*). However, we find in our data that both the appearance and mass line integrals of multiple $Na^+$ cluster types in impact mass spectra are strongly dependent upon $Na^+$ concentration (see below), which would explain the observed discrepancy.

***Effect of NaCl Concentration at Fixed Impact Velocity*** – In addition to its detection at Enceladus, both geochemical modeling and recent observations suggest that NaCl is present on Europa's surface as well and is likely a major constituent of the non-icy, presumably endogenous material ubiquitous across Europa (*34–39*). Ejecta lofted into the exosphere and sourced from these features will be analyzed by SUDA during flybys with Europa Clipper to characterize the endogenic and/or exogenic processes from which they may result. Here, we report impact mass spectra of salt-rich ice grains generated using aqueous 0.1 – 3.0 M NaCl solutions at ~3.9 km/s impact velocities (Fig. 3). In our analysis, we use the integral of the mass line for quantitative purposes, as this represents the total number of ions for a given mass and is not influenced by differences in mass resolution. The mass line integrals for characteristic $Na^+$ cluster ions from our experiments are shown in *SI Appendix*, Fig. S2. It is important to note that, based on previous experimental and modeling studies of droplet freezing in vacuum conditions (*26, 40–42*), evaporative cooling of water droplets under our experimental conditions may result in a 2.5-15% reduction in individual droplet mass upon freezing. This mass loss would therefore result in correspondingly higher ice grain solute concentrations in the ice grains than those in the solution from which the grains are sourced (*Water Droplet Freezing* in *SI Appendix*). Future studies using this method will therefore need to account for this source of uncertainty when deriving quantitative compositional information.

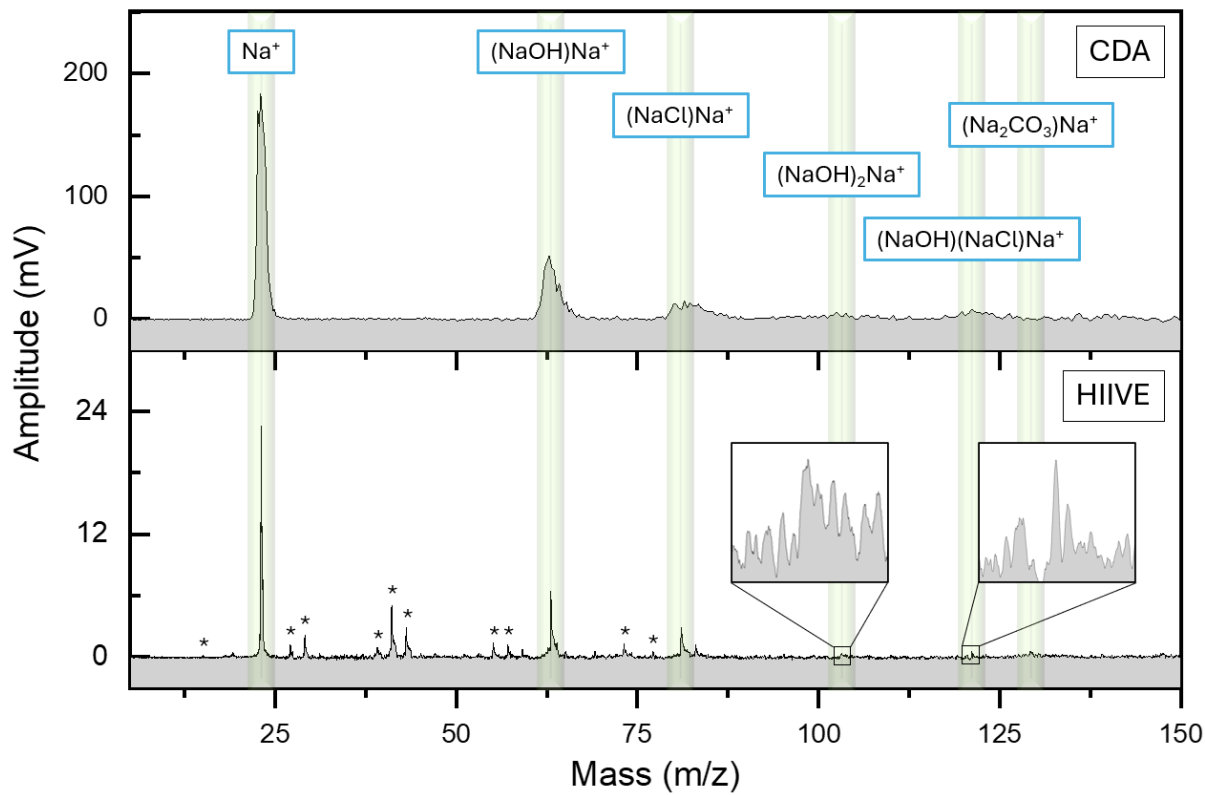

**Figure 2. Comparison of laboratory impact mass spectra with CDA data.** Mass spectral features characteristic of salt-rich E-ring ice grains analyzed by the Cosmic Dust Analyzer (CDA), which are sourced from the Enceladus plume (CDA impact event ID 180283; top), are reproduced by laboratory impact ionization mass spectra collected from ice grains generated using an aqueous solution of 0.2 M NaCl and 0.1 M $Na_2CO_3$ at 3.9 km/s impact velocities with the Hypervelocity Ice grain Impact Validation Experiment (HIIVE, bottom) in cation mode. Although this was intended to be a qualitative comparison only, these data show that our method can successfully reproduce dominant mass lines at appropriate relative ion counts for spaceborne impact mass spectra of salt-rich water ice grains. This approach therefore offers both a high-fidelity means to analyze previous data collected at Enceladus and predictive means of preparing for the Europa Clipper SUDA investigation. *Labeled peaks representing carbon contamination were confirmed through collection of a control spectrum (see *Peak Identification* in *SI Appendix*).

Our data show that, at high NaCl concentrations, cation impact mass spectra of ice grains are characterized simply by the presence of $Na^+$ and $(NaCl)_nNa^+$ clusters (unambiguously identified by integration of the $(Na^{35}Cl)_nNa^+$ and $(Na^{37}Cl)_nNa^+$ mass lines reflecting the natural abundance of chlorine isotopes), as well as minor contributions from $(NaOH)Na^+$ and $(H_2O)_{1-2}Na^+$ clusters. The peak observed at 41 m/z corresponds to a superposition of the $C_3H_5^+$ (contaminant) and $(H_2O)Na^+$ ions, which was confirmed through analysis of a control spectrum (see *Peak Identification* in *SI Appendix*). The total signal from $Na^+$ clusters was found to increase from 0.1 M to 1.0 M but is reduced from 1.0 M to 3.0 M (*SI Appendix*, Fig. S2); this is most likely an ion suppression effect, which although not widely studied in ice grain impact mass spectrometry to date, is well-documented in other mass spectrometry methods (*43*). At 0.1 M NaCl, the $(H_2O)_{1-2}Na^+$ cluster ions (41 and 59 m/z, respectively) are present; however, at 1.0 M and 3.0 M NaCl, these ions provide only a very minor contribution to the overall signal. The $(NaOH)Na^+$ cluster ion, which is a primary tracer of alkalinity, is present in the 0.1 M NaCl mass spectrum, however the signal decreases sharply at higher NaCl concentrations (*SI Appendix*, Fig. S2). This suggests that assessments of alkalinity in cation mass spectra using $(NaOH)_nNa^+$ cluster ion would be dependent upon NaCl concentration and would require this context to accurately determine using data collected at ocean worlds in the future. It is necessary to emphasize, however, that we observe these effects to be impact velocity dependent (see below).

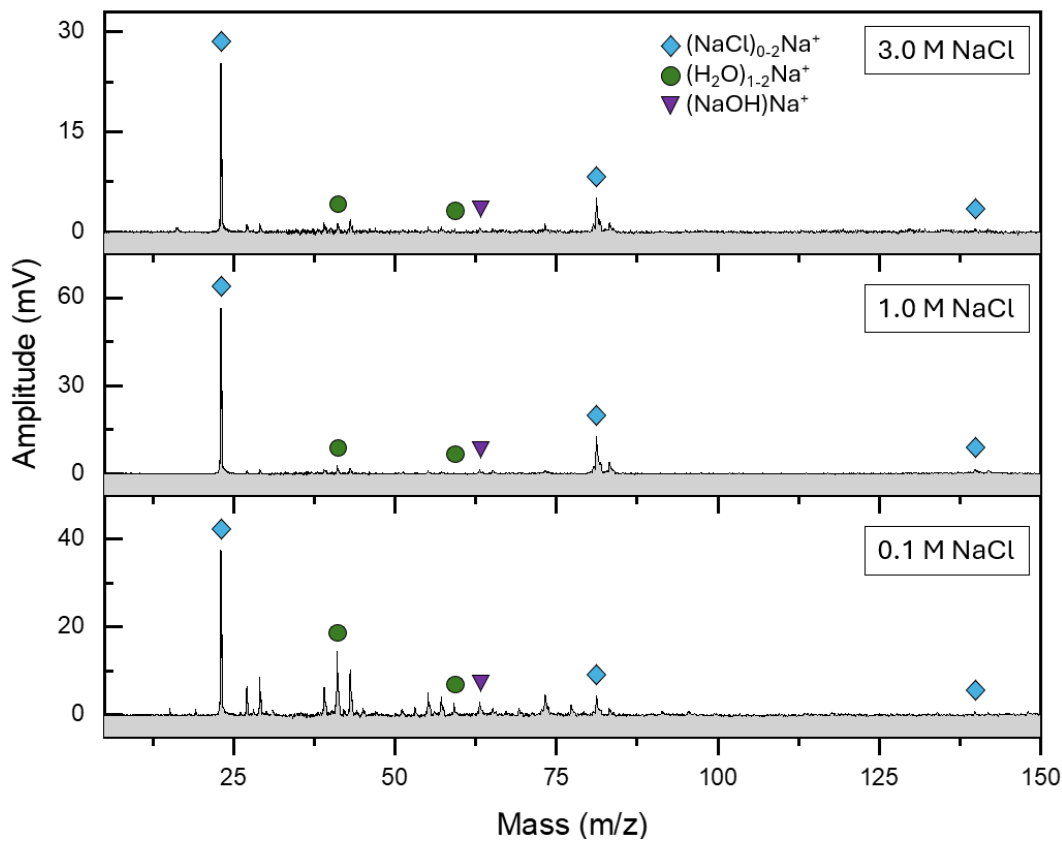

**Figure 3. Influence of salt concentration on ice grain impact mass spectra.** Impact mass spectra (3.9 km/s) collected from ice grains generated from aqueous solutions of 0.1, 1.0, and 3.0 M NaCl are characterized primarily by the $(NaCl)_{0-2}Na^+$ cluster series (diamonds); signals from the $(H_2O)_{1-2}Na^+$ (circles) and $(NaOH)Na^+$ cluster ions (triangles) provide a minor contribution to the overall signal at higher NaCl concentrations. The splitting of the $(NaCl)Na^+$ peak into masses 81 and 83 with a 3:1 ratio due to the natural abundance of $^{35}Cl$ and $^{37}Cl$ is visible. Unlabeled peaks are a result of carbon contamination confirmed through control spectra (see *Peak Identification* in *SI Appendix*).

**Dependence of Salt Clustering on Impact Velocity**

Analyses of ice grains of various compositions during the Cassini mission demonstrated that both the appearance and relative mass line integrals can be influenced by impact velocity in impact mass spectra (*12*, *44*). Although these analyses provided a largely qualitative understanding of ion formation and cluster patterns across a wide range of impact velocities, the effect of changes in impact velocity and how this influences cluster ion speciation in the range of velocities planned for future ocean world flybys is currently not well understood. Laboratory investigations examining this phenomenon have been carried out for amino acids and, to some extent, protonated water cluster ions (*25*). However, the degree to which various salt cluster signals are influenced by impact velocity and composition has not yet been examined experimentally. To study the effect of impact velocity on salt cluster speciation, experiments were conducted in which the impact velocity was varied from 2.4 – 3.6 km/s using ice grains sourced from an aqueous solution of 1.0 M NaCl (Fig. 4).

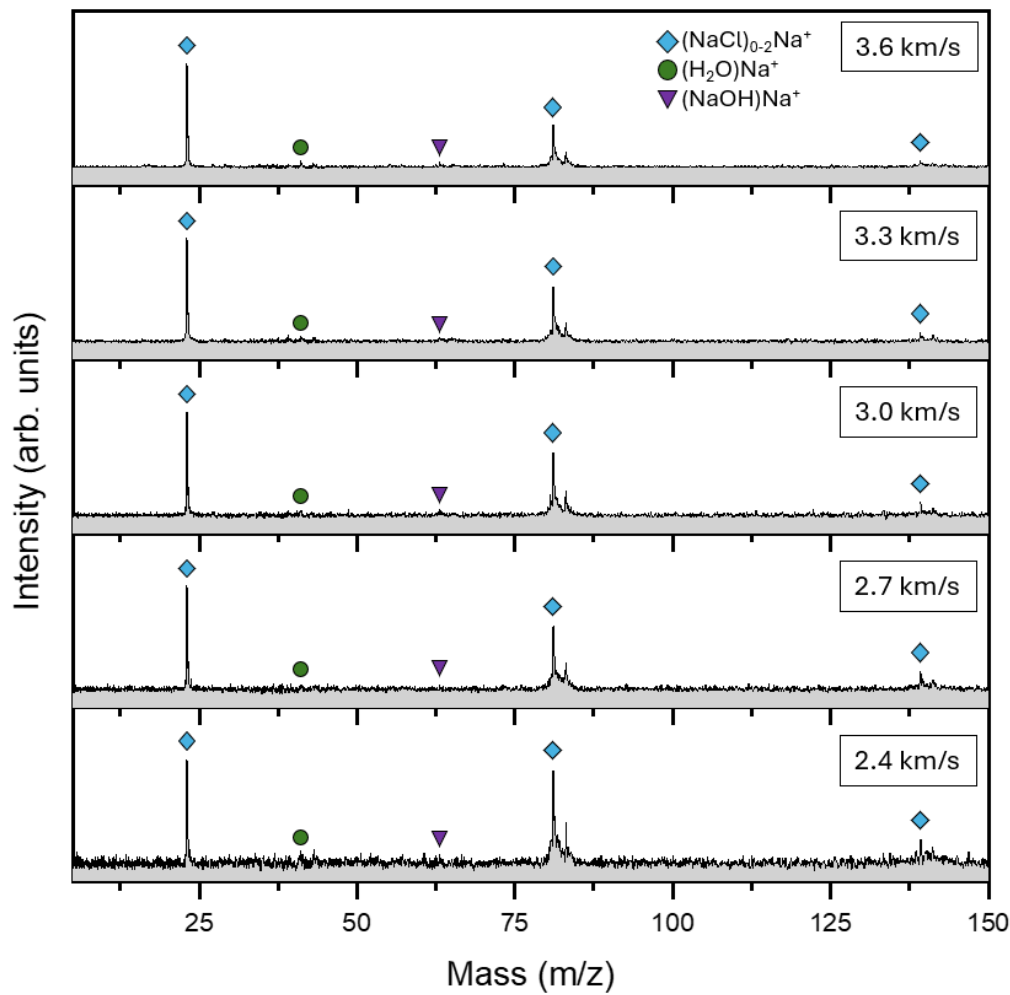

**Figure 4. Influence of impact velocity on ice grain impact mass spectra.** Salt cluster abundances in impact mass spectra collected between 2.4 – 3.6 km/s impact velocities from ice grains generated using an aqueous solution of 1.0 M NaCl suggest impact velocity can strongly cluster speciation. Mass spectra are normalized to the amplitude of the $Na^+$ ion, illustrating the degree to which integrated intensities of higher mass $(NaCl)_nNa^+$ clusters decrease relative to the $Na^+$ ion. Given that $(NaCl)_nNa^+$ mass line integrals ratios are diagnostic of NaCl concentration (see Fig. 5), impact velocity provides crucial context required for interpreting ice impact mass spectra of salt-rich ice grains.

Previous work has shown that impact ionization efficiency for sub-micron sized ice particles increases with particle velocity (*45*). Our results are consistent with this trend, as higher impact velocities (shorter extraction delay times) in our experiments yield stronger total ion signals (*SI Appendix*, Fig. S3). Although impact-velocity-dependent ion yield is likely the primary factor influencing the total ion signal, time- and velocity-dependent density variations in ice grains generated by the dispersion plume may also contribute to these signal differences to some degree.

The effect of impact velocity on organic fragmentation has been qualitatively demonstrated previously (25), however here we demonstrate that salt cluster ion signals are also influenced by relatively small changes in impact velocity. Our data show that the Na$^+$ mass line integral decreases relative to (NaCl)$_n$Na$^+$ clusters with increasing NaCl concentration, which suggests that relationships between Na$^+$ cluster sequences may be suitable for determining NaCl abundance within spaceborne ice grains (Fig. 5, left). However, a pronounced increase in the mass line integral of Na$^+$ relative to that of (NaCl)$_n$Na$^+$ was also observed with increasing impact velocity over the range of conditions examined (Fig. 5, right). Increasing NaCl concentration from 0.1 to 1.0 M reduced the Na$^+$/(NaCl)Na$^+$ line integral ratio by over a factor of two, while increasing impact velocity from 2.4 to 3.6 km/s also increased this ratio similarly by over a factor of two (Fig. 5). This suggests that velocity effects have the potential to confound assessments of salt composition in ice impact mass spectra collected by future missions if not sufficiently accounted for.

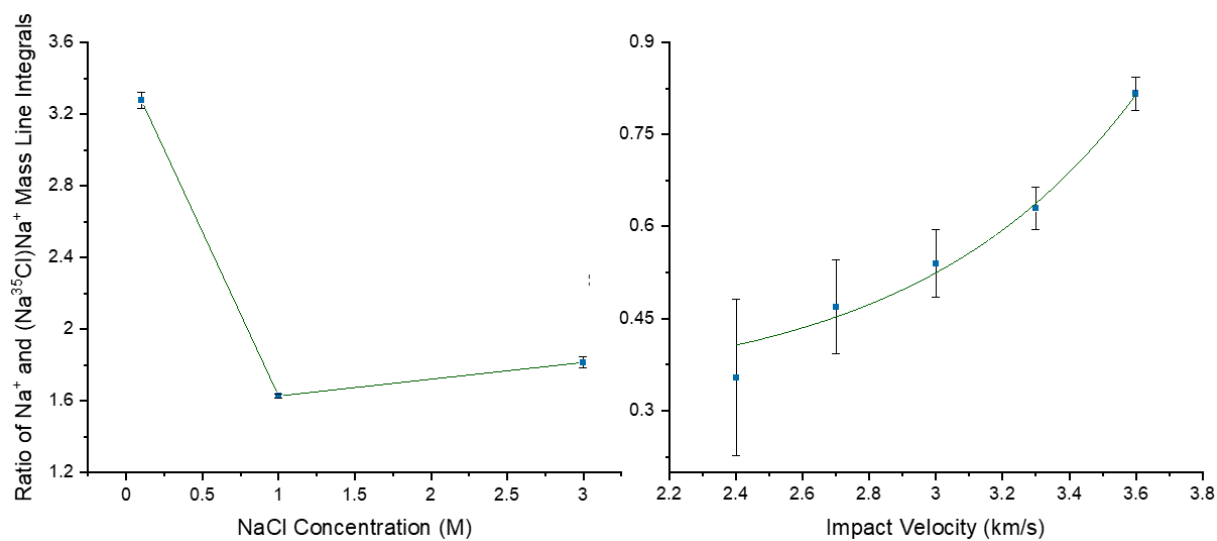

**Figure 5. Summary of concentration and velocity effects on salt-rich ice grain impact mass spectra.** The ratio of the Na$^+$ and (Na$^{35}$Cl)Na$^+$ mass line integrals is strongly influenced by both concentration (left) and impact velocity (right), emphasizing the need for this context in performing quantitative estimates of ice grain composition based on the mass line integrals of salt cluster ions (data taken from Fig. S3). Mass spectra were collected from the impacts of particles generated using aqueous solutions of 0.1, 1.0, and 3.0 M NaCl collected at 3.9 km/s (left) and an aqueous solution of 1.0 M NaCl at impact velocities ranging from 2.4 – 3.6 km/s fitted using an exponential function (right).

## Discussion

This study demonstrates the capability to experimentally replicate the flyby sampling of salty ocean world ice grains by accelerating salt-rich ice grains to spacecraft flyby velocities, impacting these particles onto a metal target representing the impact target of a spaceborne dust analyzer, and performing TOF-MS on the resulting impact plasma *in situ*. Our approach exhibits an exceptional tolerance towards hypersaline samples necessary for replicating the hypervelocity sampling of salty ocean world ice grains. These laboratory-based ground-truth experiments will be crucial for maximizing the science return of the Europa Clipper mission, as well as future missions to sample ice grains in our solar system and beyond using impact ionization mass spectrometers (e.g., dust analyzers).

By averaging signals resulting from many different impact events, our approach compensates for the signal variations characteristic of single-grain impact mass spectrometry. Single-grain impact methods typically average or co-add hundreds of individual mass spectra to generate a representative mass spectrum of a given sample (*10*, *12*, *13*, *25*); this step is achieved in parallel with our analysis method, which enables the rapid collection of individually representative impact mass spectra not only as a function of ice grain composition, but with velocity control as well. A representative mass spectrum consisting of 512 averaged mass spectra requires less than two minutes of collection time using this approach, which enables the rapid collection of datasets required to facilitate the interpretation of spaceborne impact mass spectra. Although impact ionization mass spectrometry fundamentally suffers from mass line broadening due to plasma expansion effects (*22*), our approach achieves a mass resolution of $>250$ m/$\Delta$m (FWHM) at masses $<100$ m/z, which is comparable to the mass resolution achievable by SUDA (*4*) and therefore allows for direct comparison to data acquired by the Europa Clipper mission. To our knowledge, our approach represents the highest mass resolution achieved to date through laboratory-based ice grain impact ionization mass spectrometry.

Previous analyses of Enceladus E-ring ice grain datasets suggest that the appearance and relative signal of $(H_2O)_nH^+$ clusters and the ratio of the $Na^+$ and $(H_2O)Na^+$ mass line integrals appear to be sensitive to NaCl concentration in impact mass spectra (*10*), which was also observed by Burke et al. for ice grains containing $\leq 10$ mM NaCl (*25*). We observe this effect at concentrations from 0.1 – 3.0 M NaCl as well, especially between 0.1 and 1 M NaCl, where the mass line integral of $(H_2O)Na^+$ water clusters decreases sharply with increasing NaCl concentration relative to that of $Na^+$. In doing so, we demonstrate that quantitative conclusions may be drawn from characteristic mass spectral features present in spaceborne impact mass spectra, and that cation impact mass spectra are sufficient to determine the salinity of NaCl-rich ice grains. Compositional differences between the ice grains analyzed in our Enceladus ocean simulant experiments and those sourced from the Enceladus ocean (*10*) therefore likely contribute to the relative prominence of $(H_2O)_nNa^+$ clusters in our data compared to CDA datasets collected in Saturn's E-ring (for example, the data shown in Fig. 2). This could suggest that Enceladus ice grain salinity and/or pH may be higher than the values used for our experiments, however this would require further detailed study to confirm.

Our results also emphasize the degree to which salt clustering in impact mass spectra may be sensitive to impact velocity. Our experiments suggest that relatively small changes in encounter velocity during a planetary flyby ($<1$ km/s) would influence the relative mass line integrals of cluster ion signals in the resulting mass spectrum. Given that our results suggest salt cluster ion abundances are diagnostic of salt composition and highlight their potential utility as quantitative tracers of ice grain composition (see Fig. 3), impact ionization studies capable of controlling impact velocity will provide crucial context required for accurately interpreting impact mass spectra of salt-rich ice grains. Moreover, the differences we observe in our experiments occur within the range of planned flyby velocities planned for the Europa Clipper mission, which are expected to vary across multiple km/s (*23*). If these effects are not sufficiently controlled and accounted for, they may obfuscate chemical composition inferences from the mass spectra we collect at Europa and other ocean worlds in the future using impact ionization mass spectrometers.

Although variations in analyte concentration are naturally expected to influence cluster patterns, the mechanisms underlying these variations in ice grain impact mass spectrometry are currently not well-understood due to the unique chemistry and physics of impact plasmas (*22*). The cluster speciation in the resulting mass spectrum is determined by combination of ions formed through the impact itself and those formed through the physical recombination of ions and neutrals within the impact plasma (*46*). Because of this, one would expect differences between cluster speciation in impact mass spectra and those of other ionization methods. Indeed, we find that impact mass spectra of NaCl-rich ice grains differ substantially from those obtained through analogue studies simulating the impact process through laser desorption. In addition to generating most clusters species observed in our data, previous analogue experiments used in the interpretation of spaceborne dust analyzer data and were shown to generate prominent $(NaCl)_x(H_2O)_nNa^+$ cluster sequences through the direct analysis of laser desorption ions (*47*). These analogue mass spectra suggested that many different chlorine isotopologues of higher-order $(NaCl)_x(H_2O)_nNa^+$ clusters may be present in impact mass spectra of NaCl-rich ice grains due to the natural abundance of $^{35}Cl$ and $^{37}Cl$. This would result in relatively intense and complex cluster patterns in higher-mass regions that, if present, would strongly overlap with mass lines characteristic of many different organic compounds. Klenner et al. demonstrated through analogue experiments that amino acids may form $Na^+$ adducts at the salt concentrations used in our experiments, however some of these adducts were difficult to resolve from neighboring $(NaCl)_x(H_2O)_nNa^+$ water clusters present at higher masses (*48*). Our data show that impact ionization mass spectra of NaCl-rich ice grains are characterized simply by intense $(NaCl)_nNa^+$ cluster ions (with minor contributions from $(H_2O)_nNa^+$ and $(NaOH)_nNa^+$ cluster sequences), and would therefore likely not preclude the observation of most organic molecules entrained in ice grains through the superposition of associated mass lines, including biosignature molecules such as amino acids. These results demonstrate that, if present, most organic signatures would remain distinguishable from cluster patterns arising from an NaCl-rich sample matrix and would therefore be readily identifiable in spacecraft data.

## Materials and Methods
*Instrument Design* – Laser-induced dispersion (LID) of an ultra-thin liquid water beam has traditionally been used as a soft ionization method in the analysis of biomolecules (*21*) and, more recently, as an analogue experiment simulating ice grain impacts through laser-induced ionization using high laser energies (*20*). LID under typical conditions results in the ionization of the liquid water matrix and generates a characteristic bell-shaped cluster ion distribution (*SI Appendix*, Fig. S4), however at lower laser energies (<2 mJ/pulse), this process generates predominantly nm- to μm-sized water droplets (*21*). Our ice grain acceleration and impact ionization mass spectrometry (MS) system, termed the Hypervelocity Ice grain Impact Validation Experiment (HIIVE; Fig. 6, left), uniquely utilizes LID as a method of ice grain production and acceleration to generate hypervelocity ice grain impacts. This approach exploits the high velocities at which water droplets are produced using low-energy LID and the rapid freezing the droplets undergo in high-vacuum conditions. The ice grains travel at velocities of 1.9 – 4.5 km/s towards a metal target in a high vacuum orthogonal extraction chamber, producing an impact plasma that is analyzed directly via time-of-flight (TOF)-MS using pulsed ion optics in the same manner as a dust analyzer during a planetary flyby. We observe very few to no desorption ions in the mass spectrum under these conditions, while signals resulting from impact ionization are preserved (Fig. 6, right), enabling peaks in the resulting mass spectra to be unambiguously attributed to hypervelocity impacts of ice grains. Impact velocity can be controlled by precise timing of the ion extraction pulse, allowing studies that vary not only ice grain composition, but impact velocity as well.

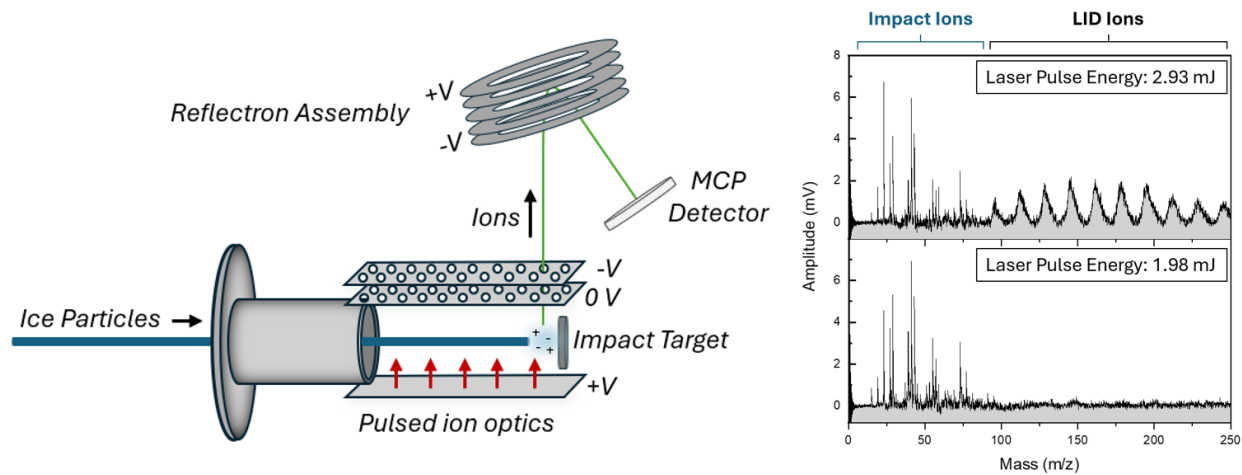

**Figure 6. Impact ionization of ice grains with HIIVE.** Ice grains generated through laser-induced dispersion (LID) of a water beam in high vacuum are accelerated towards a metal target at velocities as high as 4.5 km/s, forming an ionic plasma upon impact containing ions and neutral species reflecting the composition of the impacting material (left). Using time-delayed extraction, a voltage pulse is applied, which accelerates impact-generated ions orthogonally using dual-stage extraction and reflectron-type ion optics, generating a time-of-flight (TOF) mass spectrum in the same manner as a spaceborne impact mass spectrometer such as CDA or SUDA. Precise timing of the extraction pulse enables selection of ions within a particular impact velocity range during each experiment. Precise control over laser pulse energy inhibits the production of desorption ions through the LID process (right), allowing peaks in the mass spectrum to be unambiguously attributed to the impacts of ice grains.

*Ice Grain Production and Acceleration* – Water droplets are generated by impinging a laser (Opolette HE 2940, Opotek; 20 Hz, 5-7 ns pulse length) tuned to an -OH absorption band of water (2940 nm) onto an ultra-thin water jet generated by flowing water injected vertically at a pressure of 10-20 bar through a 30 μm diameter quartz nozzle (Advanced Microfluidic Systems GmbH) in vacuum (~$10^{-3}$ Torr). Each laser pulse causes liquid dispersion of the water jet, which generates water droplets that freeze rapidly through evaporative cooling and form ice grains as they pass from the droplet source region through a 1.5 mm molecular beam skimmer (Beam Dynamics, Inc.) and into a high vacuum (~$10^{-7}$ Torr) orthogonal ion extraction region (*Water Droplet Freezing* in *SI Appendix*). Precise control of laser pulse energy avoids the formation of desorption ions that would be generated through the LID process (*SI Appendix*, Fig. S5). Ice grains impact a circular metal target of variable composition (depending on application; in this work, a silver impact target is used) centered in the ion extraction region, producing an ionic plasma in the same manner as an ice grain impact onto the target of a spaceborne impact ionization mass spectrometer such as CDA or SUDA. The system uses a high-pressure liquid chromatography (HPLC) pump (Microliquids GmbH) to deliver samples and requires at least 1 mL of sample per analysis in its current configuration. To maintain vacuum, the water jet terminus is captured below the droplet source region and frozen using a liquid nitrogen-cooled cryotrap, with an additional cryotrap placed ahead of the turbomolecular pump to protect the pump from water vapor generated from the jet in the source region.

*Impact Mass Spectrometry* – After ice grains impact the metal target, ions from the resulting impact plasma are analyzed through time-delayed extraction with a pulse length of 2 μs defined using a digital delay generator (model DG645, Stanford Research Systems). All mass spectra were collected in positive ion mode, however reversing voltage polarities enables the analysis of negative ions using this method as well. After a variable specified time-delay corresponding to particle impact velocity, any ions present in the extraction region are drawn/repelled through an extraction grid by a pulsed DC voltage and passed through a dual-stage extraction grid, which is electrically connected to a flight tube liner. After acceleration, the ions pass through an Einzel lens where they are focused towards a dual microchannel plate detector (MCP) using a reflectron assembly (Jordan TOF Products, Inc.). A combination of dual-stage extraction and reflectron-type ion optics achieves a mass resolution of >250 m/Δm (FWHM) at masses <100 m/z. MCP voltage signals are amplified using a digital preamplifier (model 9306, Ortec), and the resulting raw waveforms are collected on an oscilloscope (model MDO34, Tektronix) in the time domain with acquisition times of approximately 40 μs. At the impact velocities observed during our experiments, the likelihood of forming multiply charged ions is negligible (*12*); therefore, we assume singly charged ions in our analyses.

*Impact Velocity Selection* - With a known distance between the water jet and the impact plate, the time-delayed pulsed ion optics used in this system allow for the selective analysis of ice grain impacts as a function of velocity by varying the time between the laser pulse and the orthogonal ion extraction pulse (*SI Appendix*, Table S1). The particle impact velocity was calculated using Eq. (1):

$$v_{particle\_impact} = \frac{d_{particle\_flight}}{t_{delay}} \tag{1}$$

Where $v_{particle\_impact}$ is the velocity of the particle impacting the target, $d_{particle\_flight}$ is the distance the particle travels from its production at the water jet to the impact plate (214.6 mm), and $t_{delay}$ is the pulsed ion extraction delay time (48 to 110 μs). The extraction pulse length results in a range of impact velocities measured by the detector as a function of extraction delay timing, which results in a combination of ions generated at the specified impact velocity and those that impact the detector within the extraction pulse window. A 2 μs extraction pulse width would therefore result in a maximum of a 180 m/s range in velocities at a 48 μs extraction delay (ions generated from impacts between 4470 – 4290 m/s; *SI Appendix*, Table S1). The range of impact velocities measured at extraction decreases with increasing extraction delay timing (and lower impact velocity). Potential uncertainties in the calculated impact velocities are attributed to uncertainty in the distance towards the impact plate resulting from variations in the nozzle position between experiments (<1 mm), as well as uncertainty in the position of impact ions upon experiencing the electric field in the orthogonal extraction region due to plasma self-shielding (*49*). Variations in nozzle positioning between experiments correspond to a maximum of ~0.5% relative uncertainty (~10-20 m/s). The maximum relative uncertainty in the position of extracted impact ions due to the distance between the impact plate and the entrance of the orthogonal extraction region (12 mm), assuming a plasma expansion velocity of ~10 km/s (*50*), is 0.625% (~15-30 m/s). We therefore assume a conservative error estimate for our velocity calculations of 50 m/s.

***Data Analysis*** – Waveforms were exported as .csv files and processed using OriginPro 2020b software. Time-of-flight spectra were converted to the mass scale empirically using characteristic peaks produced in aqueous 0.1 M solutions of NaCl and KCl including the $(H_2O)_nNa^+$, $(H_2O)_nK^+$, $(NaCl)_nNa^+$, and $(KCl)_nK^+$ cluster series. Each mass spectrum presented here represents an average of 512 individual spectra collected under identical conditions and was baseline corrected. All impact mass spectra were treated with a second-order Savitsky-Golay filter. Uncertainty in peak areas from integration of mass lines was calculated as the product of the baseline signal, which was calculated as the standard deviation of signal amplitude between 5 – 10 m/z where no analyte peaks are present, the number of data points beneath each peak, and the average spacing between m/z values beneath each peak. Uncertainty in the ratio of mass line integrals was determined through error propagation.

***Sample Preparation*** – NaCl (>99.5%) and $Na_2CO_3$ (>99.5%) were purchased from Sigma Aldrich (St. Louis, MO) and used as received. All solutions were prepared using water purified to an 18.2 MΩ-cm resistivity using a Milli-Q IQ 7005 Ultrapure and Pure Water Purification System (MilliporeSigma). Solutions were injected through a syringe filter (pore size 0.45 µm) using a Luer Lock syringe at volumes ranging from 1-10 mL per analysis.

## Acknowledgments

The authors would like to thank Bernd Abel, Ales Charvat, Robert Continetti, Jonathan Lunine, Sabrina Feldman, Greg Davis, and Mike Malaska for their technical and strategic guidance in making HIIVE possible. A portion of this work was conducted at the Jet Propulsion Laboratory, California Institute of Technology, under a contract with the National Aeronautics and Space Administration (80NM0018D0004). The authors gratefully acknowledge the NASA Postdoctoral Program, administered by Oak Ridge Associated University (ORAU), as well as internal research and technology development funding from JPL. Reference herein to any specific commercial product, process, or service by trade name, trademark, manufacturer, or otherwise, does not constitute or imply its endorsement by the United States Government or the Jet Propulsion Laboratory, California Institute of Technology. ©2025. All rights reserved.**Author contributions:**
  Conceptualization: KMS, BLH, MLC, SK
  Methodology: KMS, BLH, SEW, MECM
  Investigation: KMS, BLH
  Visualization: KMS
  Supervision: BLH, PDA, MLC
  Writing—original draft: KMS
  Writing—review & editing: All authors

**Competing interests:** Authors declare that they have no competing interests.

# Supplementary Materials for

# Replicating the flyby sampling of salty ocean world ice grains using impact ionization mass spectrometry

K. Marshall Seaton,[1]*† Bryana L. Henderson,[1]* Sascha Kempf,[2] Sarah E. Waller,[1] Morgan E. C. Miller,[1] Paul D. Asimow,[3] Morgan L. Cable[1]

[1]NASA Jet Propulsion Laboratory, California Institute of Technology, Pasadena, CA.
[2]Laboratory for Atmospheric and Space Physics, University of Colorado, Boulder, CO.
[3]Division of Geological and Planetary Sciences, California Institute of Technology, Pasadena, CA.

*Corresponding authors. Email: marshall.seaton@lasp.colorado.edu; bryana.l.henderson@jpl.nasa.gov

†Present address: Laboratory for Atmospheric and Space Physics, University of Colorado, Boulder, CO.

**This PDF file includes:**

**Supplementary Text**

Peak Identification

We observe a small amount of background carbon contamination in our mass spectra predominantly in the form of alkyl carbocations, which are observed at 15, 27, 29, 39, 41, 43, 55, 57, 73, and 77 m/z (corresponding to most likely $CH_3^+$, $C_2H_3^+$, $C_2H_5^+$, $C_3H_3^+$, $C_3H_5^+$, $C_3H_7^+$, $C_4H_7^+$, $C_4H_9^+$, $C_4H_9O^+$, and $C_6H_5^+$, respectively). To account for background levels and confirm the signals attributed to our salt solutions, we performed a background subtraction on impact mass spectra using blank measurements of ice grains formed from water purified to an 18.2 MΩ-cm resistivity. Mass spectra of both 0.1 M NaCl and the blank measurement were baseline corrected and normalized by amplitude to the $C_3H_7^+$ peak at 43 m/z. The blank measurement was then subtracted from the 0.1 M NaCl mass spectrum (Fig. S1). Our experiments showed that the addition of NaCl to the solution enhanced the impact-induced fragmentation of carbon impurities, which results in more intense carbon fragments in NaCl solutions, however ion suppression effects at higher NaCl concentrations (~1.0 M and above) reduces organic signals by a factor of >6. These experiments confirm $Na^+$, $(H_2O)Na^+$, $(H_2O)_2Na^+$, $(NaOH)Na^+$, and $(NaCl)Na^+$ $(NaCl)_2Na^+$ cluster ions as the primary species observed in NaCl-rich ice grain impact mass spectra. We note that the peak observed at 41 m/z in our data is likely a superposition of the $C_3H_5^+$ and $(H_2O)Na^+$ ions; however, we demonstrate that background subtraction can be used to deconvolve the relative contributions of these species. The peak identified as an alkyl carbocation fragment at 77 m/z could, in principle, be due to signal contribution from the $(H_2O)_3Na^+$ ion. However, control experiments conducted using aqueous CsCl solutions (data not shown) contained the 77 m/z mass line while the preceding $(H_2O)_2Na^+$ cluster at 59 m/z was absent, leading to the conclusion that the 77 m/z mass line is due to an organic fragment. Although the peak at 39 m/z could, in principle, be attributed to $K^+$, its relative abundance in terrestrial groundwater suggests this is unlikely given their relative signals in our background measurements (*51*). Therefore, the signal contribution to the peak at 39 m/z was attributed primarily to the $C_3H_3^+$ fragment.

Water Droplet Freezing

Previous theoretical and experimental studies of the dispersion process suggest that droplets generated through LID (2940 nm laser, 5-7 ns pulse length, ~2 mJ pulse energy) range from nm- to μm-sized (*21, 52*). Conveniently, this size distribution falls within the Cassini-derived distribution of ice particle sizes within the Enceladus plume (*53*) and may also be consistent with particles in Europa's plumes, if present (*54*). Extrapolation of models studying the freezing of larger water droplets indicates that water droplets within this expected size range rapidly freeze within tens of μs through evaporative cooling in the high vacuum environment used in HIIVE experiments (~$10^{-3}$ and $10^{-7}$ Torr in the source and ion extraction regions, respectively) which is within the timescales required for LID-generated water particles to reach the impact target in our experiments (*55–59*). Miller et al. (*60*) extended these modeling studies to incorporate the multi-phase freezing and supercooling of water droplets in similar vacuum conditions to those used here and showed that a 950 nm droplet freezes rapidly and crystallizes within ~30 μs, forming an ice grain. This model suggests a mass loss of ~2.5% through evaporative cooling, however several other models have suggested that droplets shrink by up to 5-15% by volume as they freeze through evaporative cooling, assuming volume nucleation as the primary freezing mechanism (*40–42*). This would result in a slight increase in concentration of solute ions or molecules during the freezing process, which will need to be accounted for when performing quantitative studies using this technique in the future.

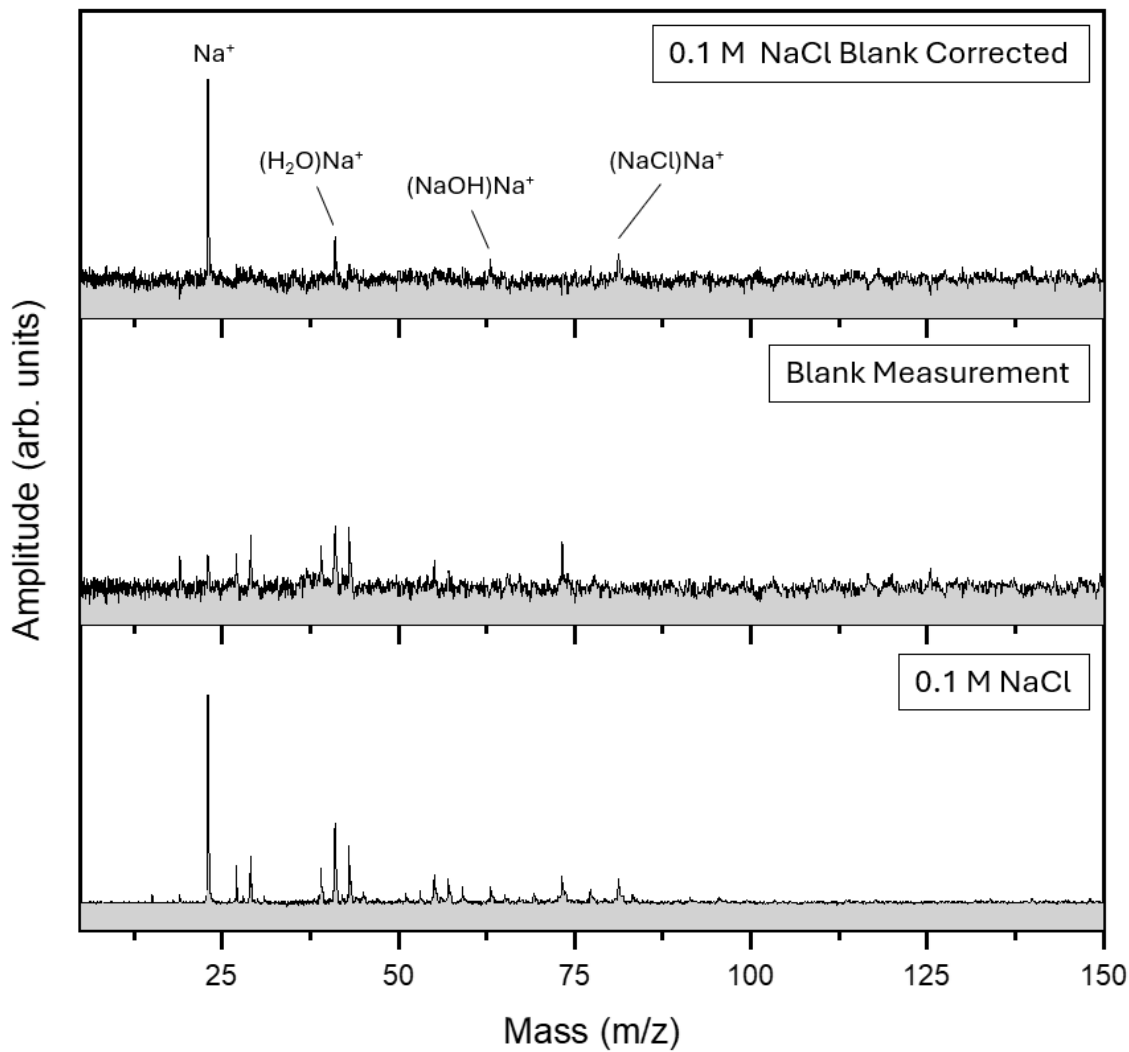

**Fig. S1.**
Background subtraction of a 0.1 M NaCl solution confirms the presence of $Na^+$, $(H_2O)Na^+$, $(NaOH)Na^+$, and $(NaCl)_nNa^+$ cluster ions in NaCl-rich ice impact mass spectra. Blank measurements were performed using water purified to an 18.2 MΩ-cm resistivity. All mass spectra were normalized to the amplitude of the mass line at 43 m/z and collected at an impact velocity of 3.9 km/s.

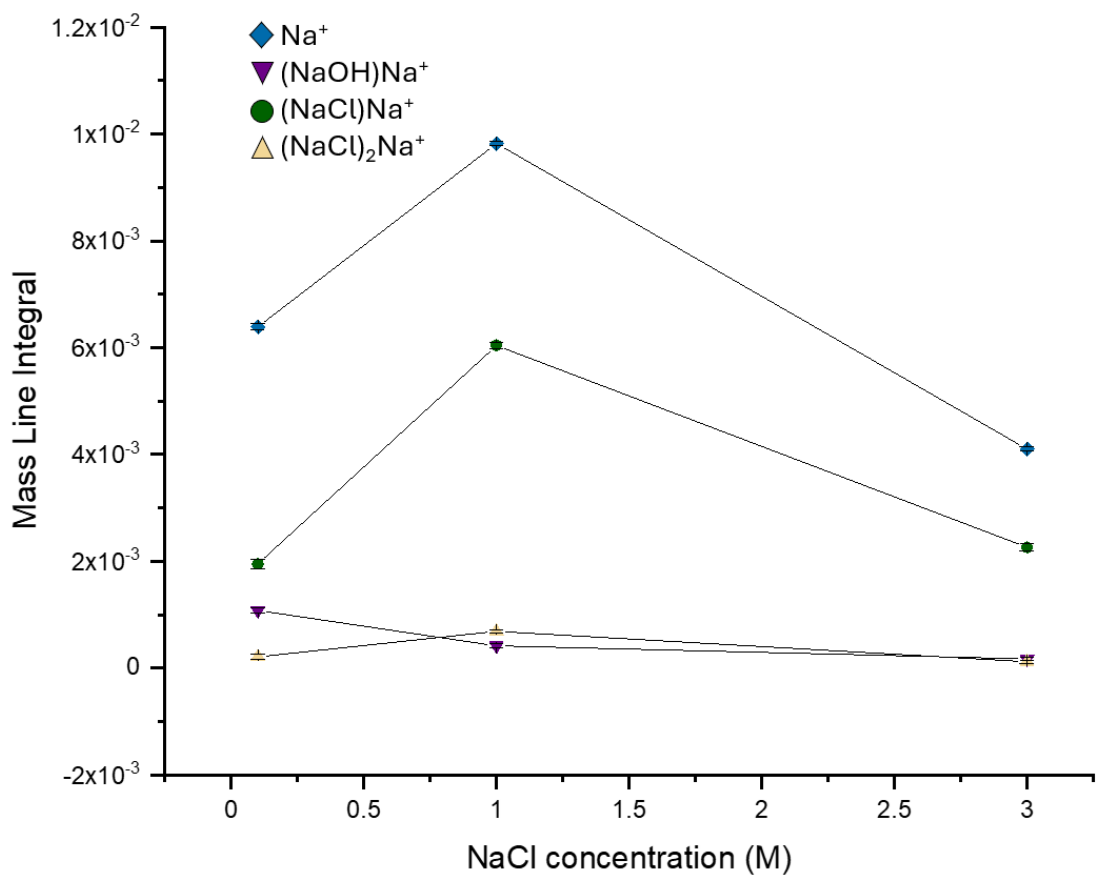

**Fig. S2.**
The total integrated intensities of the $Na^+$, $(NaOH)Na^+$, $(NaCl)Na^+$, and $(NaCl)_2Na^+$ cluster ions are strongly dependent on NaCl concentration in ice impact mass spectrometry (data taken from Fig. 3). Mass spectra were collected from 3.9 km/s impacts of ice grains generated using aqueous NaCl solutions at 0.1, 1.0, and 3.0 M.

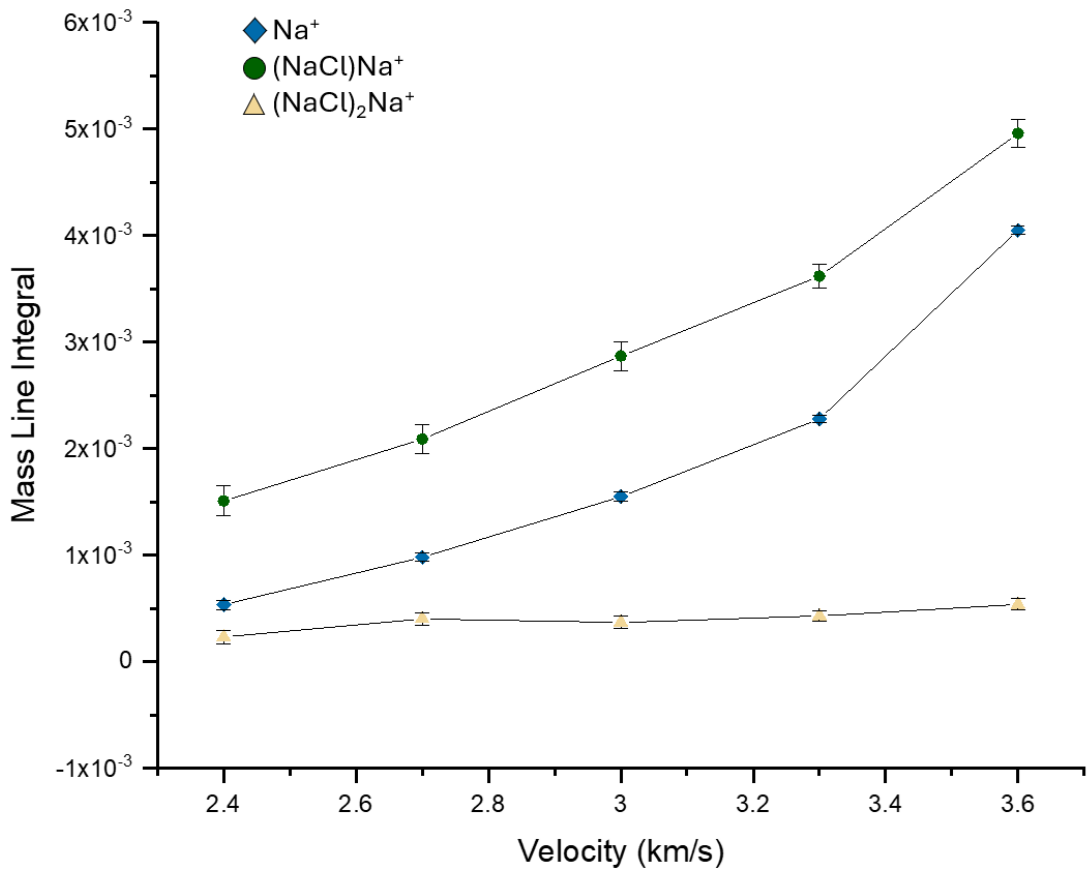

**Fig. S3.**

Mass line integrals of the $(NaCl)_{0-2}Na^+$ cluster series increase with increasing particle impact velocity (data taken from Fig. 4). Mass spectra were collected from the impacts of particles generated using an aqueous solution of 1.0 M NaCl at impact velocities ranging from 2.4 – 3.6 km/s.

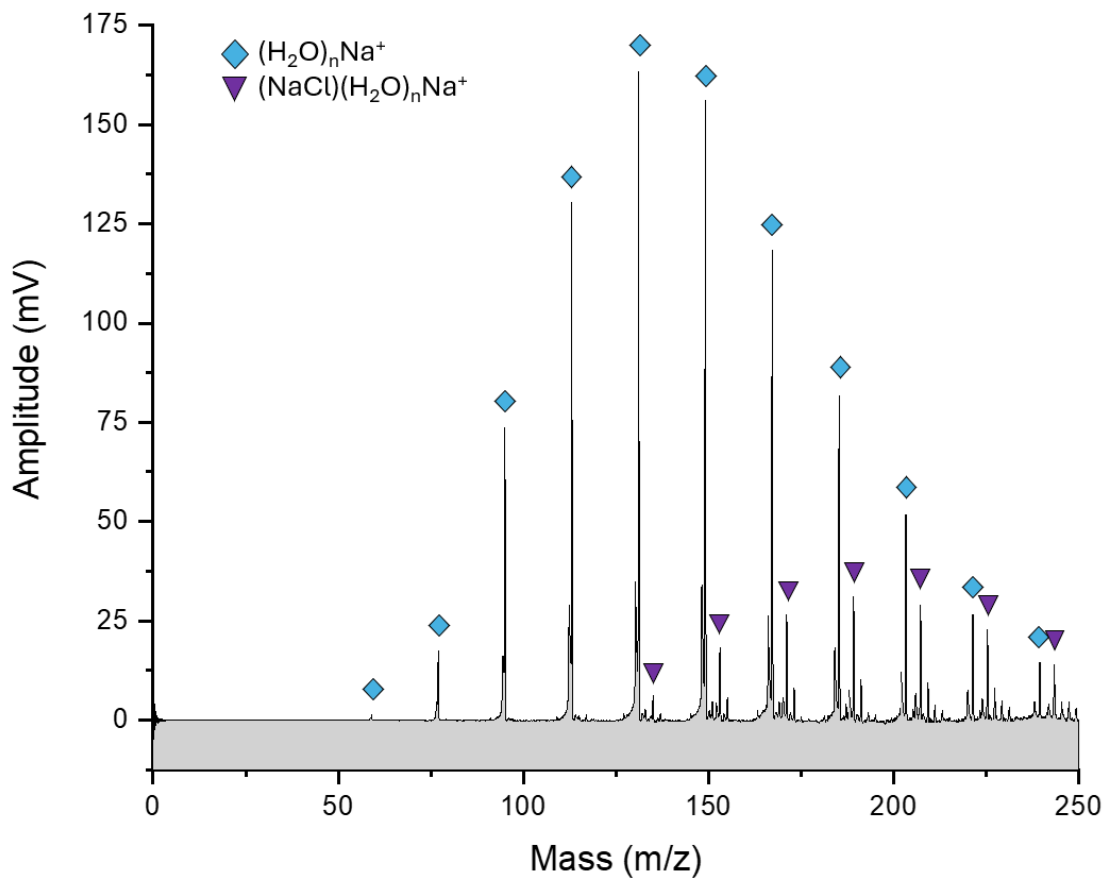

**Fig. S4.**
Mass spectra from the laser-induced dispersion (LID) of an aqueous 0.1 M NaCl solution collected in the absence of an impact target at 55 μs extraction delay time (corresponding to a 3.9 km/s theoretical impact velocity) using a pulse energy of 2.93 mJ. Ions generated through LID show a characteristic bell-shaped distribution, with no signals observed for masses below 59 m/z at the laser energies used here (<3.0 mJ per pulse). In addition to the presence of higher-mass $Na^+$ water clusters, the mass spectrum is dominated by $(NaCl)(H_2O)_nNa^+$ clusters as well, which are characteristic of LID mass spectra using concentrated aqueous NaCl solutions (*47*). This contrasts with impact ionization mass spectra of NaCl-rich solutions, which are characterized by a prominent $Na^+$ ion, an absence of water clusters at high salt concentrations, and a gradual decay in peak intensity as a function of mass (see Fig. 3).

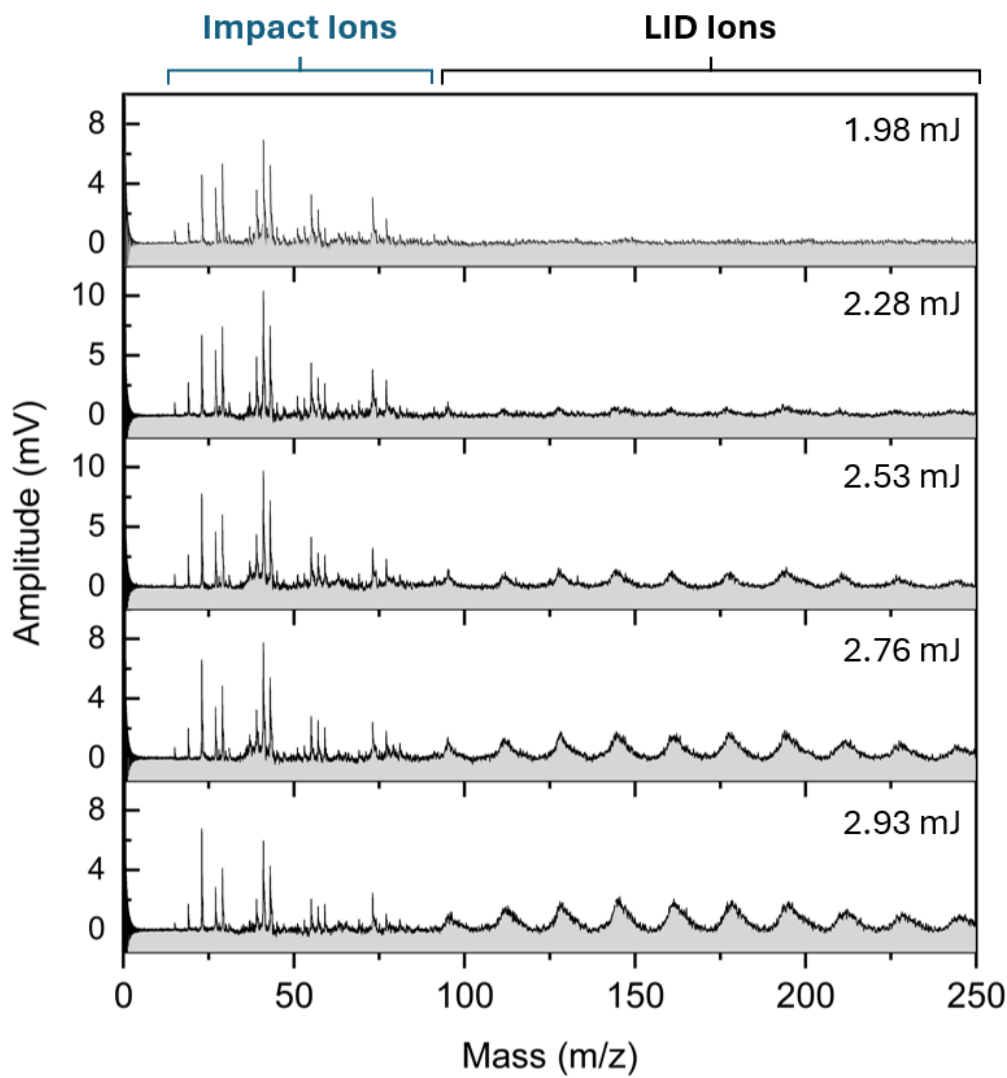

**Fig. S5.**
Reducing laser pulse energy from 2.93 to 1.98 mJ (measured 5 cm from the laser output and adjusted using a variable Q-switch delay) eliminates ions generated through laser-induced dispersion (LID) of an aqueous 10 mM NaCl solution while preserving impact ion signals. Ions generated through hypervelocity impact are much higher in energy (~5-10 eV; (*50*)) than those generated through LID at the point of ion extraction (*61*); these energy differences, in addition to spatial focusing, results in poor resolution of LID signals (broad features roughly present between 100-250 m/z) under the focusing conditions under which impact signals (sharp features present roughly below 100 m/z) are clearly resolved. Because of this, any potential LID ions appearing in our impact mass spectra have a characteristic mass resolution that differs significantly from the mass resolution of impact ions for a given mass, and therefore can be easily identified and controlled for.

**Table S1.**

Relationship between extraction delay timing and the range of particle impact velocities extracted under each condition.

| Extraction Delay (μs) | Impact velocity (m/s) |
|---|---|
| 48 | 4470 - 4290 |
| 50 | 4290 - 4130 |
| 52 | 4130 - 3970 |
| 54 | 3970 - 3830 |
| 56 | 3830 - 3700 |
| 58 | 3700 - 3580 |
| 60 | 3580 - 3460 |
| 62 | 3460 - 3350 |
| 64 | 3350 - 3250 |
| 66 | 3250 - 3160 |
| 68 | 3160 - 3070 |
| 70 | 3070 - 2980 |
| 72 | 2980 - 2900 |
| 74 | 2900 - 2820 |
| 76 | 2820 - 2750 |
| 78 | 2750 - 2680 |
| 80 | 2680 - 2620 |
| 82 | 2620 - 2560 |
| 84 | 2560 - 2500 |
| 86 | 2500 - 2440 |
| 88 | 2440 - 2380 |
| 90 | 2380 - 2330 |
| 92 | 2330 - 2280 |
| 94 | 2280 - 2240 |
| 96 | 2240 – 2190 |
| 98 | 2190 - 2150 |
| 100 | 2150 - 2100 |
| 102 | 2100 - 2060 |
| 104 | 2060 - 2030 |
| 106 | 2030 - 1990 |
| 108 | 1990 - 1950 |
| 110 | 1950 - 1920 |